\documentclass[12pt]{JHEP3}

\usepackage{epsfig,multicol,amsmath}
\newcommand\fverb{\setbox\pippobox=\hbox\bgroup\verb}
\newcommand\fverbdo{\egroup\medskip\noindent%
\fbox{\unhbox\pippobox}\ }			
\newcommand\fverbit{\egroup\item[\fbox{\unhbox\pippobox}]}
\newbox\pippobox
\def\d2bar{$\overline{\mbox D2}$}
\title{Non-supersymmetric Attractor with the Cosmological Constant}
\author{Jin-Ho Cho$^{1,2}$ and Soonkeon Nam$^{1}$\\
$^{1}$Department of Physics \& Research Institute for Basic Sciences,\\ Kyung Hee University, Seoul 130-701, Korea\\
$^{2}$Center for Quantum Space Time, Sogang University, Seoul 121-742, Korea\\\\
E-mail: \email{cho.jinho@gmail.com}, \email{nam@khu.ac.kr}}

\abstract{As a test for the non-supersymmetric attractor mechanism, we consider extremal Reissner-Nordstr\"{o}m-(anti-)de Sitter black holes. Based on the simple observation that the near-horizon geometry of a generic extremal black hole contains two-dimensional anti-de Sitter factor even in the presence of the positive cosmological constant, we apply Ashoke Sen's entropy function method to compute the entropy of these  black holes. We find the results which exactly agree with the Bekenstein-Hawking entropy. We also obtain the constant higher-order correction to the entropy due to the Gauss-Bonnet term.}  


\begin{document} 
\section{Introduction}
Supersymmetry plays an important role in our accounting for the statistical origin of the extremal black holes\cite{Strominger:1996sh}. In string theory, the degrees of freedom pertaining to the black hole entropy in the strong coupling regime is nothing but those of the strings living on D-branes, that is,  the would-be black hole in the weak coupling regime. Supersymmetry protects the number of the degrees of freedom living on D-branes from disappearing under the variation of the string coupling.

The extremal black holes  in $(3+1)$-dimensional $\mathcal{N}=2$ supergravity exhibit the attractor behavior\cite{Ferrara:1995ih,Strominger:1996kf,Ferrara:1996um}, one of whose consequence is that the entropy concerns the values of the vector moduli just on the horizon and does not care about their asymptotic values. In view of the attractor equation derivable from the Killing spinor equation (for example, concerned with the gaugino variation in IIB case), it looks like that the role of the supersymmetry preserved by the extremal black holes is still important here.  

Meanwhile, Sen showed that it is only the near horizon geometry that is necessary 
in determining the entropy of extremal black holes  and we do not need any information about supersymmetry\cite{Sen:2005iz}. This suggests that the attractor behavior is quite generic in every extremal black hole that has anti-de Sitter (AdS) geometry near its horizon. There have been a lot of works\cite{Goldstein:2005hq,Tripathy:2005qp,Alishahiha:2006ke,Kallosh:2006bt,Chandrasekhar:2006kx,Kaura:2006mv,Chandrasekhar:2006ic} about non-supersymmetric attractor. In most of these works, non-BPS but extremal black holes in the background of various supersymmetric vacua were considered.

A stringent test for the non-supersymmetric extremal attractor would be to check the entropy function of the extremal black holes  embedded in non-supersymmetric vacua such as de Sitter spacetime. In this paper, we consider the Reissner-Nordstr\"{o}m-de Sitter (RNdS) black hole, which is obviously non-supersymmetric by construction but has extremal analogues. We also consider the Reissner-Nordstr\"{o}m-anti-de Sitter (RNAdS) case. The attractor with the negative cosmological constant was discussed in Ref. \cite{Morales:2006gm}, though in the framework of supersymmetric theory, that is, the gauged supergravity.  

In Sec. \ref{secii}, we discuss general properties of the black holes and show that the near horizon geometry of a generic extremal black hole includes two-dimensional AdS space-time even in de Sitter background. In Sec. \ref{seciii}, we discuss various extremal cases of RN(A)dS black holes. Since the nature of RNdS black hole varies with its size, we refine our interests to the cold black hole. In Sec. \ref{seciv}, we show that for a given value of the charge $Q$, the Bekenstein-Hawking entropy is an increasing function of the $4$-dimensional cosmological constant $\Lambda_4$. We compute the entropy function of the RN(A)dS black hole based on the near-horizon geometry discussed in Sec. \ref{secii}. The extremal value of the entropy function shows a perfect agreement with Bekenstein-Hawking entropy. Sec. \ref{secv} discusses various points of our results including a constant contribution of Gauss-Bonnet term to the black hole entropy in $4$-dimensions. 

\section{The Event Horizon of a Static Black Hole}\label{secii}
A static black hole is specified by a single function $f(r)$ in the metric, that has at least one zero. In the conventional static black hole of the form,
\begin{equation}\label{qq}
ds^{2}=-f(r)dt^{2}+ \frac{1}{f(r)}dr^{2}+ r^{2}d\Omega^{2}_{d-2}, 
\end{equation}
the function $f(r)$ encodes all the properties of a static black hole like the event horizon, Hawking temperature, or the entropy. At each zero $r_{0}$ of the function $f(r)$, the hypersurface $r=r_{0}$ becomes null and the coordinate time $t$ will be infinitely red-shifted, which characterize the event horizon of the static black hole.

\subsection{The Temperature of a Static Black Hole}\label{seciia}
Hawking temperature is always positive both in the region interior to and exterior to the event horizon as long as the coordinate $r$ is spatial. To see this, we just use the Euclidean argument of determining Hawking temperature. The main point of the argument is that the event horizon is just a coordinate singularity. There is nothing special about the event horizon for the freely falling observer. 

In the Euclidean version, the approximate near horizon geometry looks like just a plane times a sphere. Assuming a zero of $f(r)$ at $r=r_{0}$, one can define the near horizon coordinate as 
\begin{equation}\label{q3}
r-r_{0}\equiv \pm \epsilon\rho.
\end{equation}  
The upper sign is for the region exterior to the horizon and the lower one for the region interior to the horizon so that the near horizon coordinate $\rho$ is alwasy positive. Expanding the function $f(r)$ in the dimensionless parameter $\epsilon$ as
\begin{equation}\label{taylor1}
f(r)=\pm f'(r_{0})\epsilon \rho + \mathcal{O}(\epsilon^{2}),
\end{equation}  
we get the metric near the horizon;
\begin{equation}\label{q4}
ds^{2}\simeq\pm \epsilon \left(-f'(r_{0}) \frac{\rho^{2}}{4}dt^{2}+ \frac{1}{f'(r_{0})} d\rho^{2}\right) +r^{2}_{0}d\Omega^{2}_{d-2}.
\end{equation}
The prime in the function $f(r)$ stands for the derivative with respect to the coordinate $r$.  
Since we are interested in the region where $f(r)$ is positive so that the signature of the temporal coordinate $t$ and the spatial coordinate $r$ be not switched, $f'(r_{0})$ is positive in the exterior region and negative in the interior region. In other words, we are interested in the region where $\pm f'(r_{0})$ is positive.  

The slope of the function $f(r)$ concerns Hawking temperature. By taking Wick rotation as $\lambda=i t$, we get the metric describing a plane times a sphere;
\begin{equation}\label{q21}
ds^{2}_{E}\simeq \frac{\epsilon}{\vert f'(r_{0})\vert} \left(\vert f'(r_{0})\vert^{2} \frac{\rho^{2}}{4}d\lambda^{2}+ d\rho^{2}\right) +r^{2}_{0}d\Omega^{2}_{d-2}.
\end{equation}
For generic period of Euclidean time $\lambda$, the metric has a conical singularity at the horizon $\rho=0$. Since the horizon is not special to the free falling observer, this sort of the singularity should be absent, which requires the period be $\triangle \lambda= 4\pi/\vert f'(r_{0})\vert$ \cite{Susskind:1994sm}.
The Hawking temperature $T$ is defined as the inverse of the period of Euclidean time; 
\begin{equation}\label{q5}
T= \frac{1}{\triangle}=\frac{\vert f'(r_{0})\vert}{4\pi}.
\end{equation} 

\subsection{Near Horizon Geometry of the Extremal Black Holes}\label{seciib}

The extremal black holes are the ones with coincident outer and inner horizon. Mathematically this happens when the function $f(r)$ has a double zero, that is, not only $f(r_{0})=0$ but also $f'(r_{0})=0$. The near horizon geometry of the extremal black holes are always factorized as a two-dimensional  anti-de Sitter space-time and a sphere. This is valid even in de Sitter background. The expansion of the function $f(r)$ goes up to the next order in $\epsilon$;
\begin{equation}\label{taylor2}
f(r)=\frac{f''(r_{0})}{2}(\epsilon \rho)^{2} + \mathcal{O}(\epsilon^{3}).
\end{equation} 
In the near horizon coordinates $\pm\epsilon\rho=r-r_{0}$ and $\tilde{t}/\epsilon=t$, the metric can be approximated as 
\begin{eqnarray}\label{q6}
ds^{2}&\simeq&-\frac{1}{2}f''(r_{0})\rho^{2}\tilde{t}^{2}+ \frac{2}{f''(r_{0})} \frac{d\rho^{2}}{\rho^{2}}+ r^{2}_{0}d\Omega^{2}_{d-2},
\end{eqnarray}
which describes the factorization, AdS$_2\times$S$^{d-2}$. Note that $f''(r_{0})$ is positive because we are interested in the region where $f(r)>0$. In general, the size of the anti-de Sitter space-time,
\begin{equation}\label{q7}
l^{2}_{{\text ads}_{2}}= \frac{2}{f''(r_{0})}
\end{equation}  
need not be equal to the size $r_{0}$ of the sphere\footnote{The feature of AdS$_2$ near horizon geometry of the extremal black holes persists even in $(1+1)$-dimensions. There, the entropy concerns the value of the dilaton field at the horizon. See Ref. \cite{Hyun:2007ii} for details.}.

\section{The Reissner-Nordstr\"{o}m-(anti-)de Sitter Black Holes}\label{seciii}

In this paper, we concern the extremal black holes in the presence of the cosmological constant. Let us look into the basic properties of $4$-dimensional RN(A)dS black holes. Since the properties are very sensitive to the size of the black hole, we refine our interests to the cold black hole by specifying the range of the size. The details, about the classification of the black holes for non-extremal cases, can be found in Refs. \cite{Romans:1991nq,Cai:1997ih,Cai:2001tv}.

\subsection{Various Horizons}\label{iiia}

The geometry of charged black holes in $d$-dimensional (anti-)de Sitter background is as follows:
\begin{eqnarray}\label{q8}
ds^{2}&=&-f(r)dt^{2}+ \frac{1}{f(r)}dr^{2}+ r^{2}d\Omega^{2}_{d-2}, \nonumber\\
f(r)&=&1- \frac{\omega_{d-2}M}{r^{d-3}}+ \frac{(d-2)\omega^{2}_{d-2}Q^{2}}{8(d-3)r^{2(d-3)}}-\frac{\eta}{l^{2}}r^{2},\qquad \omega_{d-2}= \frac{16\pi G_{d}}{(d-2)\mbox{Vol}(S^{d-2})},
\end{eqnarray}
where $G_{d}$ is $d$-dimensional Newton's constant and $l$ is the length scale characterizing the cosmological constant, that is,
\begin{equation}\label{q9}
\frac{\eta}{l^{2}}= \frac{2}{\left(d-1 \right) \left(d-2 \right) }\Lambda_{d},\qquad\quad (d\ge 3)
\end{equation}  
with $\eta= +1/-\!1$ for dS/AdS respectively. The charge parameter $Q$ (that has the dimension of inverse length) is related to the electric charge $q$ and the magnetic charge $p$ as
\begin{eqnarray}\label{pqcharge}
F^{(2)}_{e}&=&\frac{q}{r^{d-2}\mbox{Vol}(S^{d-2}) }\, dt\wedge dr,\qquad F^{(d-2)}_{m}=\frac{p}{\mbox{Vol}(S^{d-2}) }\,d\Omega_{d-2},\nonumber\\
4\pi G_{d} Q^{2}&=& q^{2}+p^{2}.
\end{eqnarray}

In general, $f(r)$ can have up to $2(d-2)$ zeros, corresponding to the horizons, of which we are usually interested in the largest three values $a<b<c$. In $4$-dimensions, the positions, $r=a,\,b,$ and $c$ correspond to the inner, the outer, and the cosmological horizon respectively. The other zero of $f(r)$ in $4$-dimensions is negative, thus unphysical.

In this paper, we focus on the $4$-dimensional case only. This is the simplest nontrivial case for the black hole entropy. Higher dimensional cases, though look straightforward, involve a bit messy expressions for the event horizon and the results are expected to be different from $4$-dimensional case. Those cases will be discussed elsewhere.

\subsection{The Extremal Cases}\label{iiib}
Let us first consider the case when $f(r)$ has a double zero at $r=b$ and a simple zero at $r=c$. Then it can be rewritten as
\begin{equation}\label{factor}
f(r)=- \frac{\eta}{l^{2}r^{2}} \left(r-b \right)^{2} \left(r-c \right) \left( r+ \left(2b+c \right) \right). 
\end{equation}  
For this specific factorization, the zeros are related with the parameters $M,\,Q$ and $l$ as
\begin{eqnarray}
&&\frac{3\eta}{l^{2}}b^{4}-b^{2}+G^{2}_{4}Q^{2}=0,\label{20a}\\
&&\frac{2\eta}{l^{2}}b^{3}-b+G_{4}M=0,\label{20b} \\
&&\frac{\eta}{l^{2}}\left( 3b^{2}+2bc+c^{2}\right)-1=0.\label{20c} 
\end{eqnarray}
The size of the event horizon can be read from (\ref{20a});
\begin{equation}\label{20.9}
b^{2}= \frac{l^{2}}{6\eta}\left(1\pm \sqrt{1- \frac{12\eta}{l^{2}}G^{2}_{4}Q^{2}} \right). 
\end{equation}  

The possible range of the double root $b$ depends on the value of $\eta=\pm 1$. In  anti-de Sitter case, that is when $\eta=-1$, there is no restriction on $b$. From Eqs. (\ref{20a}) and (\ref{20b}), we see that a positive double zero point $b$ can be always found for any given positive value of $G_{4}M$ and of $G^{2}_{4}Q^{2}$. (See the left figures in Fig. \ref{Gmass} and Fig. \ref{Gcharge}.) It is determined by (\ref{20.9}) with the lower choice of the sign. The upper sign results in an imaginary value of $b$.

\FIGURE{
\epsfig{file=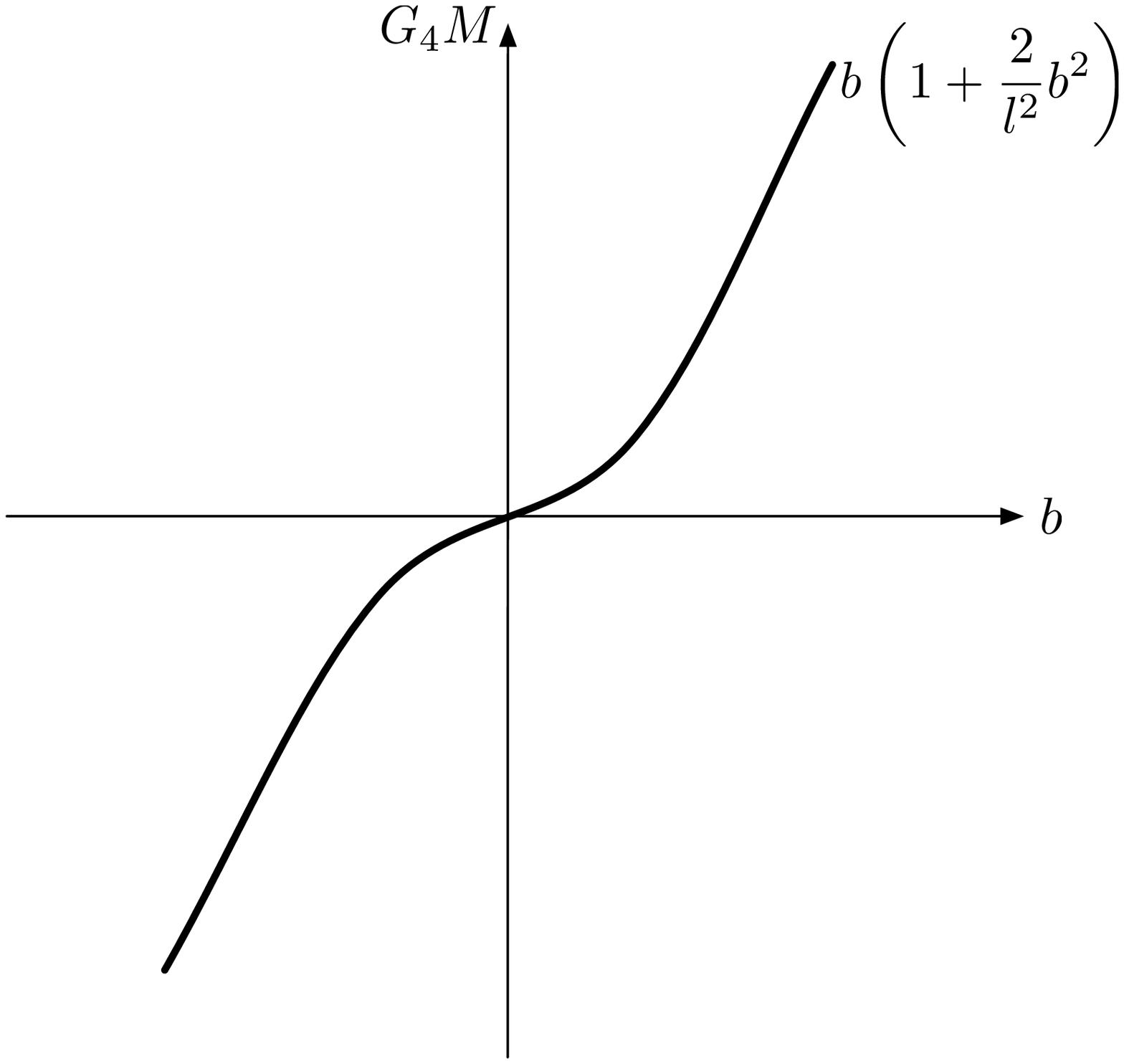,width=7cm}
\epsfig{file=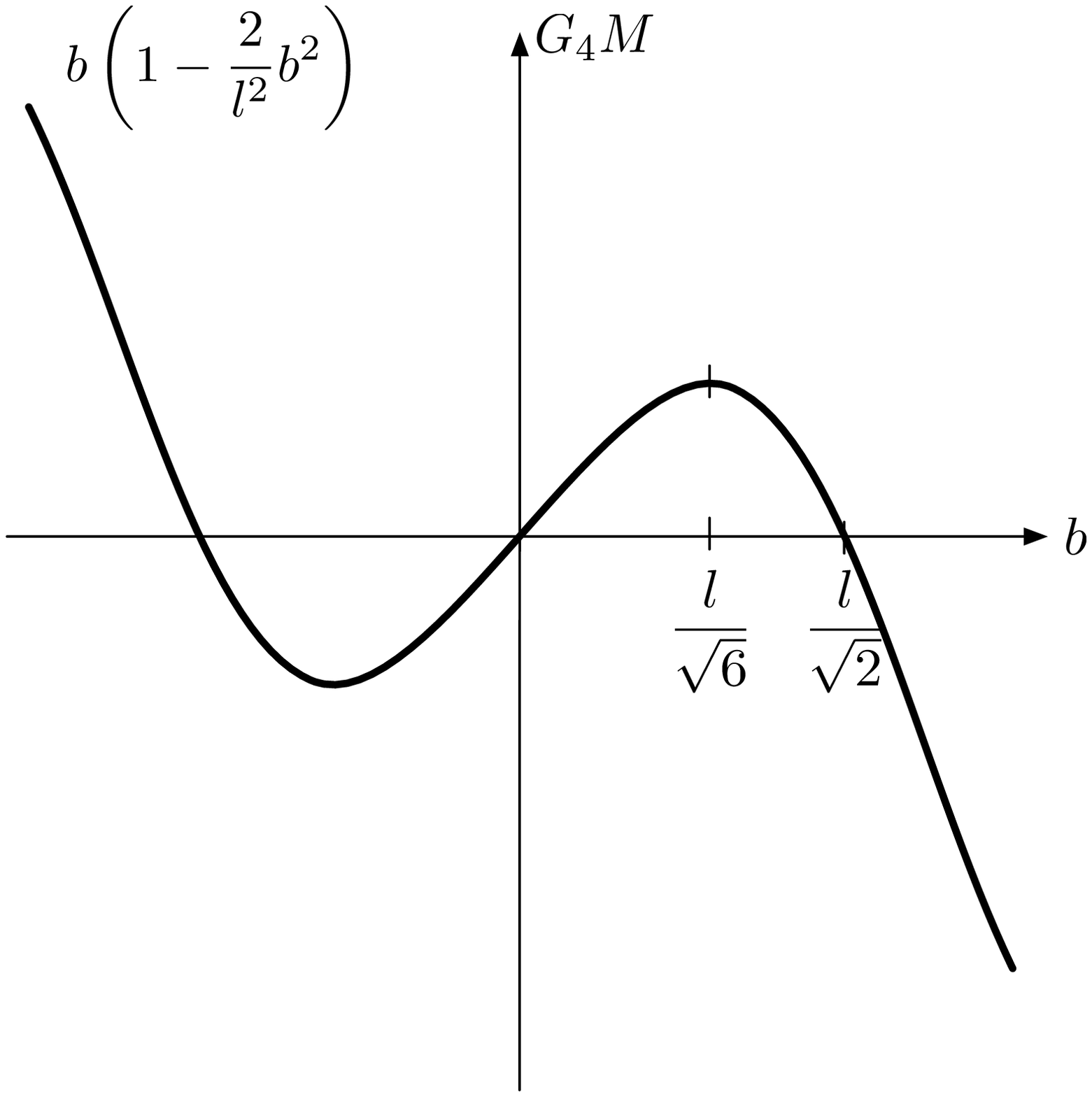,width=7cm} 
        \caption{Masses as functions of the double zero $b$. The left/right figure is for the anti-de Sitter/de Sitter case.}
	\label{Gmass}
}


Meanwhile, in de Sitter case (the right figures in Fig. \ref{Gmass} and Fig. \ref{Gcharge}), that is when $\eta=1$, the double zero $b$ should be less than $l/\sqrt{2}$ to ensure a non-negative value of $G_{4}M$. The maximum value of $G_{4}M$ is $2l/3\sqrt{6}$ at $b=l/\sqrt{6}$. Moreover, in Eq. (\ref{20a}), we note that a non-negative value of $G^{2}_{4}Q^{2}$ is possible only when $b$ is less than $l/\sqrt{3}$. The point $b=l/\sqrt{3}$ corresponds to the Nariai black hole, that is, the extremal Schwarzchild-de Sitter black hole. $G^{2}_{4}Q^{2}$ has the maximum value $l^{2}/12$ at $b=l/\sqrt{6}$. In all, the double zero should be confined as
\begin{equation}\label{bdomain}
0\leq b\leq \frac{l}{\sqrt{3}}
\end{equation}
to make the values of $G_{4}M$ and $G^{2}_{4}Q^{2}$ positive.

Another zero $c$ exists only for de Sitter case. From Eq. (\ref{20c}) we get
\begin{equation}\label{20.11}
c=-b\pm\sqrt{-2b^{2}+\eta l^{2}}
\end{equation}  
that becomes imaginary for anti-de Sitter case ($\eta=-1$), implying the absence of the cosmological horizon. In de Sitter case, the horizon position $c$ takes positive value for the range of (\ref{bdomain}) if we take the upper sign in the above result (\ref{20.11}). The lower sign corresponds to the other horizon, $-2b-c$, of the function $f(r)$ in (\ref{factor}). 

\FIGURE{
\epsfig{file=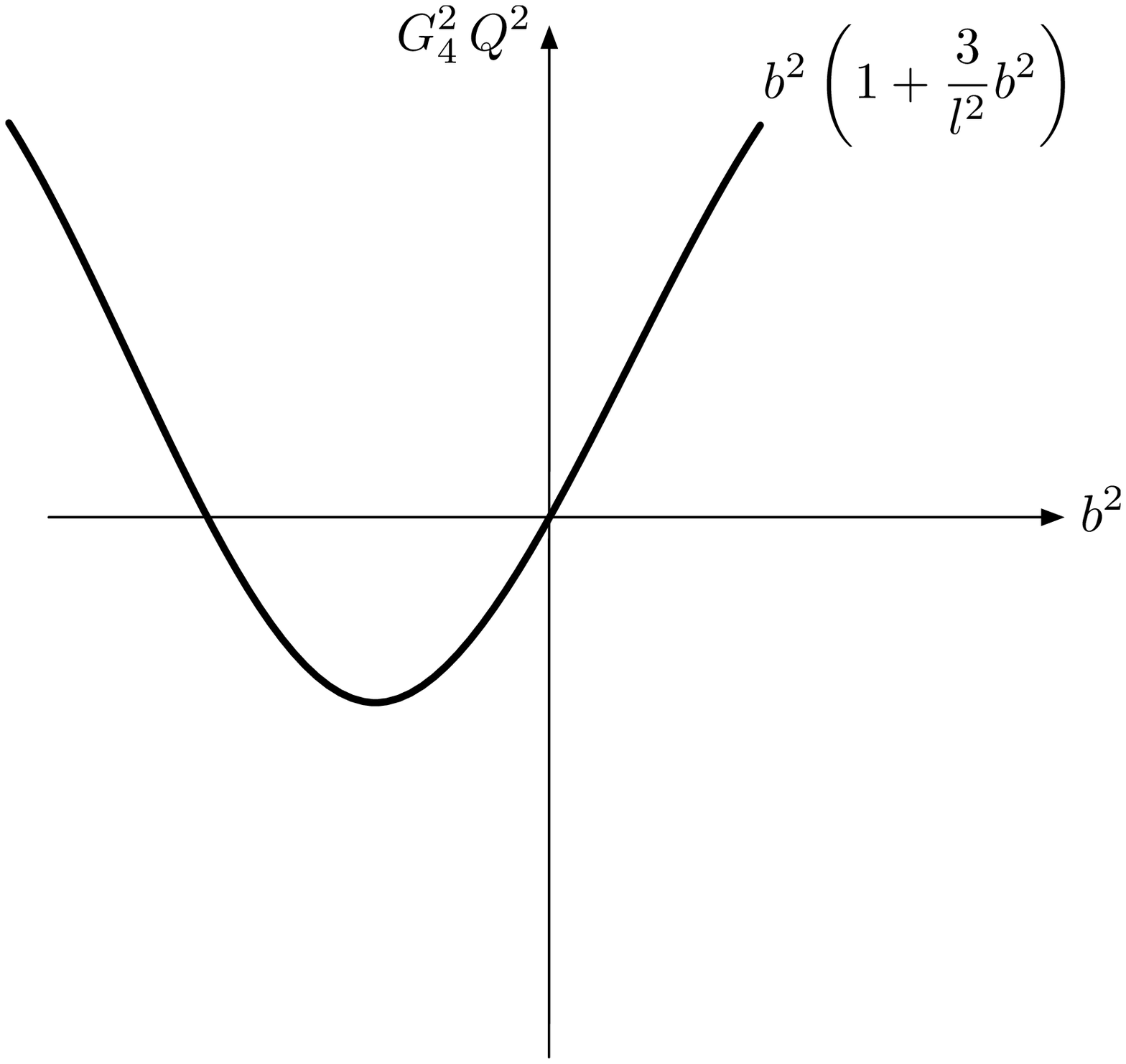,width=7cm}
\epsfig{file=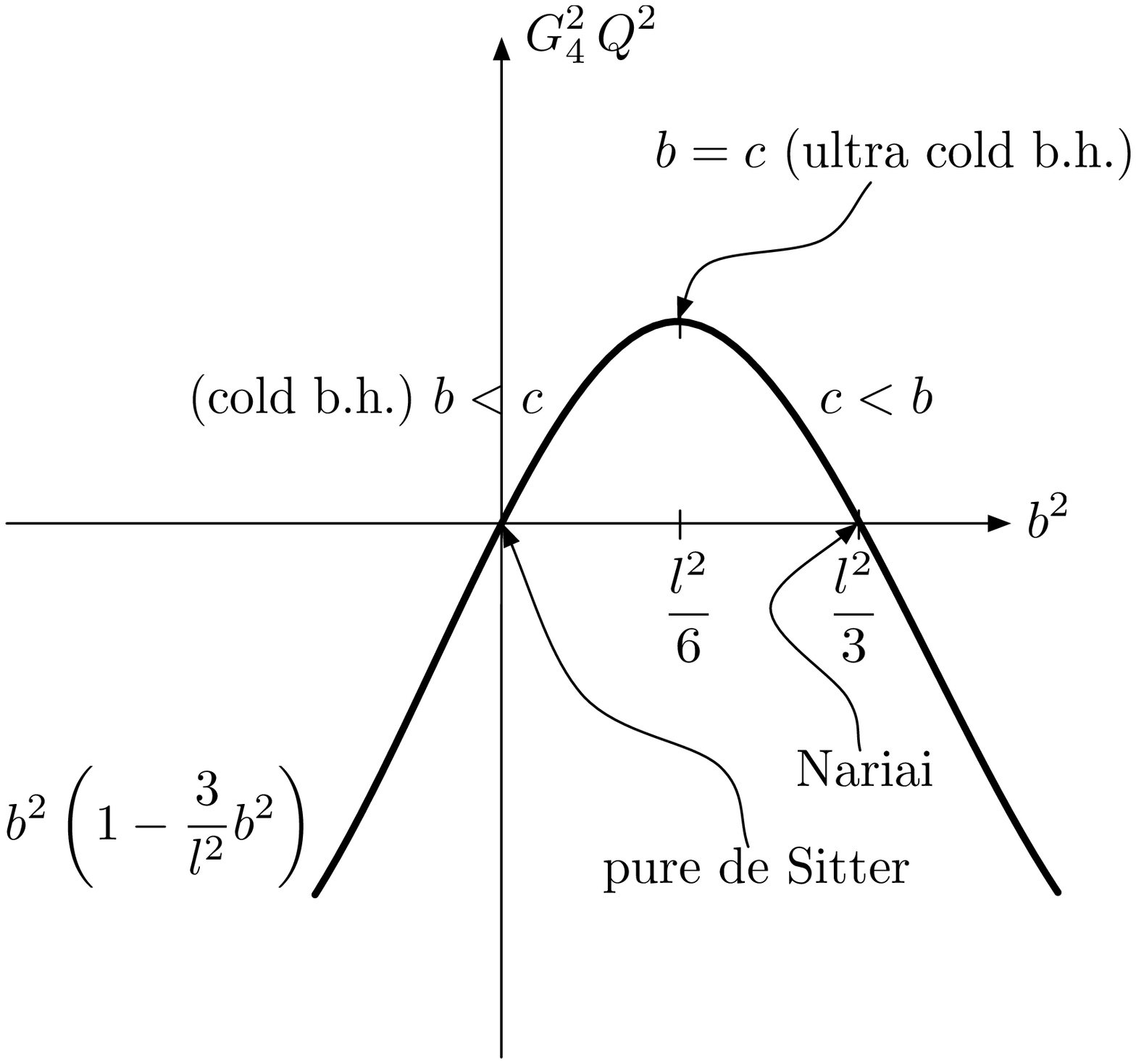,width=7cm} 
        \caption{Charges as functions of the double zero $b^{2}$. The left/right figure is for the anti-de Sitter/de Sitter case.}
	\label{Gcharge}
}


In de Sitter case, the multi-sign in Eq. (\ref{20.9}) corresponds to the relative magnitude of $b$ and $c$. In fact, making use of Eq. (\ref{20.11}), one can see that $(b-c)$ is positive when $l/\sqrt{6}<b\leq l/\sqrt{3}$, that is when the upper sign is taken in (\ref{20.9}). Especially when $G^{2}_{4}Q^{2}=l^{2}/12$, the double zero $b$ and the simple zero $c$ coincide composing a triple zero. This case is expecially called as an ultra-cold black hole because of its vanishing Hawking temperatures for both horizons at $b$ and $c$. However in this paper, we are not interested in the case of $b\geq c$ because then the function $f(r)$ is negative in the region near the double zero $b$. Therefore, we refine our interests to the case of the cold black hole (with the Hawking temperatures $T_b=0$ and $T_c\ne 0$) that is when $0<b^2<l^2/6$.

\subsection{The Cold Black Hole}\label{iiid}
In anti-de Sitter case, there exists only one positive double zero $b$ of the function $f(r)$, which corresponds to the event horizon of the extremal anti-de Sitter-Reissner-Nordstr\"{o}m black hole. The simple zero $c$ corresponding to the cosmological horizon is imaginary valued, thus no cosmological constant exists. In de Sitter case, Hawking temperature of the black holes for the case of $c>b$ is not unique. Being of the form $T_{r_{0}}=|f'(r_{0})|/4\pi$, it is zero as for the horizon at $r=b$, while non-vanishing for the horizon (cosmological) at $r=c$;
\begin{equation}\label{q10}
T_{b}=0,\qquad T_{c}=\frac{\left(c+b \right) \left(c-b \right)^{2} }{2\pi c^{2}l^{2}}.
\end{equation}  
This case is termed as a `cold black hole'.  In the special case of $b=c$, that is, an ultra-cold black hole, both temperatures $T_{b}$ and $T_{c}$ vanish.
   
\section{The Entropy Function of the RN(A)dS Black Holes}\label{seciv}

\subsection{The Bekenstein-Hawking Entropy}\label{iiie}
In this section, we derive the explicit form of the Bekenstein-Hawking entropy for the RN(A)dS black holes. 
The Bekenstein-Hawking entropy is given by the area of the horizon; 
\begin{eqnarray}\label{bhentropy}
S_{BH}&=& \frac{A}{4G_{4}}= \frac{\pi b^{2}}{G_{4}}=\frac{\pi l^{2}}{6\eta G_{4}}\left(1- \sqrt{1- \frac{12\eta}{l^{2}}G^{2}_{4}Q^{2}} \right)\nonumber\\
&=&\frac{\pi}{2G_{4}\Lambda_{4}}\left(1-\sqrt{1-4\Lambda_{4}G^{2}_{4}Q^{2}} \right) .
\end{eqnarray}  
Some properties about the result are in order. First, we note that the entropy is positive irrespective of the value of $\eta$. The entropy is an increasing function with $\Lambda_{4}$; increases from the value zero at $\Lambda_{4} \rightarrow -\infty$ to the value $2\pi G_{4}Q^{2}$ at $\Lambda_{4}=1/(4G^{2}_{4}Q^{2})$. It is also continuous at $\Lambda_{4}=0$ approaching the value,  
$S_{BH}\rightarrow \pi G_{4} Q^{2}$,
that is the entropy of a Reissner-Nordstr\"{o}m black hole in the flat background.

Second, the absence of the parameter $M$ in the above expression is due to the extremal nature of the black holes. Although there is no supersymmetry, the extremality gives some relation between the mass $M$ and the charge $Q$ like that of BPS relation.

Indeed one can obtain the relation among the parameters $Q$, $M$, and $l$ by inserting Eq. (\ref{20.9}) into Eq. (\ref{20b}). The result is
\begin{eqnarray}\label{bpslike}
9M^{2}-4Q^{2}-3M\sqrt{9M^{2}-8Q^{2}}&=&\frac{l^{2}}{3\eta G^{2}_{4}}\left(1-\sqrt{1- \frac{12\eta}{l^{2}}G^{2}_{4}Q^{2}} \right).
\end{eqnarray}
In the limit of $l \rightarrow\infty$, the right hand side limits to $2Q^{2}$, which implies the BPS relation, $Q^{2}=M^{2}$, of the extremal Reissner-Nordstr\"{o}m black holes in the flat background ($\Lambda_{4}=0$). Thanks to the relation (\ref{bpslike}), the entropy of the extremal black hole can be written in terms of the charge $Q$ only.

\subsection{The Near Horizon Geometry}\label{iiic}

The near horizon geometry around the double zero $b$ is factorized as AdS$_2\times$S$^{2}$ where the size $l_{{\text ads}_{2}}$ of the anti-de Sitter space-time and the radius $l_{{\text s}_{2}}$ of the sphere are respectively given by
\begin{equation}\label{sizeofads}
l^{2}_{{\text ads}_{2}}=\frac{2}{f''(r_{0})}=\frac{l^{2}b^{2}}{\eta\left(c-b \right) \left(c+3b \right)}= \frac{l^{2}b^{2}}{l^{2}-6\eta b^{2}}, \qquad\quad l^{2}_{{\text s}_{2}}=b^{2}.
\end{equation}   
In anti-de Sitter case, that is when $\eta=-1$, the value of $l^{2}_{{\text ads}_{2}}$ increases with $b^{2}$, approaching the maximal value $l^{2}/6$ as $b^{2}\rightarrow \infty$. In other words, however large the size of S$^{2}$ is, that of AdS$_2$ is restricted to be finite.

In de Sitter case, that is when $\eta=1$, we only consider the region $0<b^{2}<l^{2}/6$, where $c>b$ and $f(r)$ is positive. As the size (squared), $b^{2}$, of the sphere approaches the value $l^{2}/6$, that of AdS$_2$ becomes divergent. This is the opposite situation to the anti-de Sitter case. Indeed one can invert the relation (\ref{sizeofads}) into the form;
\begin{equation}\label{q11}
l^{2}_{{\text ads}_{2}}\equiv\hat{b}^{2},\qquad l^{2}_{{\text s}_{2}}= \frac{l^{2}\hat{b}^{2}}{l^{2}+6\eta\hat{b}^{2}}.
\end{equation}    
Therefore as the size (squared), $\hat{b}^{2}$ of the AdS$_2$ approaches the value $l^{2}/6$, the size of the sphere becomes divergent.

\subsection{The Entropy Function}
The entropy function is the expression for the Wald's entropy formula \cite{Wald:1993nt,Jacobson:1993vj,Iyer:1994ys,Jacobson:1994qe} applied to the near horizon geometry of some specific type of black holes. In its derivation, all the fields compatible with the isometry of the black holes simplify the expression significantly and lead to the result in the form of some Legendre transform derivable from Lagrangian density. See \cite{Sen:2005iz} for details.

Let us start with the action describing the gravitational field coupled to the electric and the magnetic fields in $d$-dimensional (anti-)de Sitter background.
Since we are now interested in the black holes (rather than higher dimensional black objects), it will be
\begin{equation}\label{actionRND}
\mathcal{S}=\int d^{d}x\sqrt{-g} \left[ \frac{1}{2\kappa}\left(R- 2\Lambda_{d} \right)- \frac{1}{4}\vert F^{(2)}_{e}\vert^{2}- \frac{1}{2\cdot (d-2)!}\vert F^{(d-2)}_{m}\vert^{2}  \right], 
\end{equation}
where $2\kappa=16\pi G_{d}$. 

Despite the presence of the cosmological constant, one may define the extremal black holes in (anti-)de Sitter space-time as the ones whose near-horizon geometries are factorized as 
\begin{eqnarray}\label{ansatz1}
ds^{2}&=&v_{1}ds^{2}_{\mbox{ads}_{2}}+v_{2}ds^{2}_{\mbox{s}^{d-2}} \nonumber\\
F^{(2)}_{e}&=&e\, dt\wedge dr,\qquad F^{(d-2)}_{m}=\frac{p}{\mbox{Vol}(S^{d-2})}\,d\Omega_{d-2}.
\end{eqnarray}
This is based on the observation, made in Sec. \ref{secii}, about the properties of an extremal black hole. In the above simple form, we redefined the temporal coordinate $t$ making use of the dimensionful parameters like $G_{d}$ so that it carry the dimension of the inverse length. From hereon, the Greek indices $\{\mu,\nu,\rho,\sigma,\cdots\}$ are pertaining to AdS$_{2}$, while the Roman indices $\{a,b,c,d,\cdots\}$ are used for the sphere S$^{d-2}$. The capital Roman indices are reserved for the whole space. For example,
\begin{equation}\label{q12}
\{x^{M}\}=\{x^{\mu};\,x^{a}\}.
\end{equation}  

Regarding the ansatz of the geometry in (\ref{ansatz1}), we have
\begin{equation}\label{riemann}
R_{ \mu\nu \rho\sigma}=-\frac{1}{v_{1}}\left(g_{\mu\rho}g_{\nu\sigma}-g_{\mu\sigma}g_{\nu\rho} \right),\qquad R_{abcd}= \frac{1}{v_{2}}\left( g_{ac}g_{bd}-g_{ad}g_{bc}\right). 
\end{equation}    
Therefore
\begin{equation}\label{riccitensor}
R_{\mu\nu}=- \frac{1}{v_{1}}g_{\mu\nu},\quad R_{ab}= \frac{1}{v_{2}}\left(d-3 \right)g_{ab}  
\end{equation}  
and 
\begin{equation}\label{scalarcurv}
R=g^{MN}R_{MN}=g^{ \mu\nu}R_{ \mu\nu}+g^{ab}R_{ab}=- \frac{2}{v_{1}}+ \frac{\left(d-3 \right) \left(d-2 \right) }{v_{2}}.
\end{equation}

The entropy function is defined as the Legendre transform of some function $L$ with the replacement of the `velocity' $e$ with its `canonical conjugate momentum' $q$ as
\begin{equation}\label{q13}
F=2\pi \left( q e -L\right).
\end{equation}  
The function $L$ is the Lagrangian density over two-dimensional anti-de Sitter space-time;
\begin{eqnarray}\label{q14}
L&=&\int d\Omega_{d-2}\sqrt{-g}\,\mathcal{L}\\
&=&\text{Vol}(S^{d-2}) v_{1}v_{2}^{\frac{d-2}{2}}\left\{\frac{1}{2\kappa}\left(- \frac{2}{v_{1}} + \frac{2}{v_{2}}-2\Lambda_{d}\right) + \frac{1}{2}\left( \frac{e^{2}}{v^{2}_{1}}- \frac{p^{2}}{\text{Vol}^{2}(S^{d-2}) v_{2}^{d-2}} \right) \right\}. \nonumber
\end{eqnarray}
It was obtained by inserting the above ansatz (\ref{ansatz1}) about the near-horizon configuration into the action (\ref{actionRND}).  

The extremum value of the entropy function gives the black hole entropy.
Again we return to $d=4$ case for simplicity. The relation between $e$ and $p$ is obtained by extremizing the entropy function $F$ with respect to $e$;
\begin{equation}\label{q15}
\frac{\partial F}{\partial e}=0\qquad \Rightarrow \qquad e= \frac{q v_{1}}{4\pi v_{2}}.
\end{equation}  
The `Hamiltonian' $F$ is then written in terms of $q,\,p,\,v_{1},\,v_{2},\,\Lambda_{d}$;
\begin{equation}\label{q16}
F= \frac{v_{1}}{4 v_{2}} \left(p^{2}+q^{2} \right)+\frac{8\pi^{2} }{\kappa}\left(-v_{1}+v_{2}+v_{1}v_{2}\Lambda_{d} \right).  
\end{equation}  

The entropy function is extremized at 
\begin{equation}\label{v1v2}
v_{1}=\frac{32\pi^{2}v^{2}_{2}}{\left(p^{2}+q^{2} \right)\kappa-32\pi^{2}v^{2}_{2}\Lambda_{4}},\qquad v_{2}=\frac{1}{8\pi\Lambda_{4}}\left(4\pi\pm \sqrt{16\pi^{2}-2\left(p^{2}+q^{2} \right)\kappa\,\Lambda_{4} } \right). 
\end{equation}
Out of two choices for the sign in $v_{2}$ of Eq. (\ref{v1v2}), we exclude the upper one because it gives negative value on $v_{1}$, the size of the anti-de Sitter space-time.  
Therefore the extremal value of the entropy function is
\begin{equation}\label{q17}
F= \frac{\pi l^{2}}{6\eta \,G_{4}}\left(1-\sqrt{1-\frac{3\eta G_{4}}{\pi l^{2}} \left(p^{2}+q^{2} \right)}  \right),
\end{equation}  
which coincides with Bekenstein-Hawking entropy of the extremal RNAdS black hole (for $\eta=-1$) or the cold black hole (for $\eta=1$);
\begin{equation}\label{q18}
S_{BH}= \frac{4\pi b^{2}}{4G_{4}},
\end{equation}  
if we set
\begin{equation}\label{q19}
G_{4}Q^{2}= \frac{1}{4\pi}\left(p^{2}+q^{2} \right). 
\end{equation}  
This latter equation is nothing but the last equation of Eq. (\ref{pqcharge}). Actually, the size $v_2$ of the sphere (in (\ref{v1v2})) extremizing the entropy function coincides with the size $b^2$ of RN(A)dS black hole obtained in (\ref{20.9}).

\section{Discussions}\label{secv}

\subsection{Universality of the Near Horizon AdS$_2$ geometry in the Extremal Black Holes}
The near horizon geometry of the extremal Reissner-Nordstr\"{o}m-(anti-)de Sitter black holes contains the anti-de Sitter space-time irrespective of the sign of the background cosmological constant. Mathematically this specific factorization of the geometry is easy to understand. Be the background cosmological constant positive or negative, the anti-de Sitter part and the sphere part adjust each of their scalar curvatures to compose that of (anti-)de Sitter background. 
This means that the sizes of the anti-de Sitter part and of the sphere part need not match each other. The curvature scalar is the sum of that of AdS part and that of the sphere part;
\begin{equation}\label{q20}
R=R_{ads_{2}}+R_{s_{2}}=- \frac{2}{l^{2}_{ads_{2}}}+ \frac{2}{l^{2}_{s_{2}}}= \frac{12\eta}{l^{2}}.
\end{equation}
In de Sitter background ($\eta=1$), the size of the AdS part is larger than that of the sphere part, that is, $l^{2}_{ads_{2}}>l^{2}_{s_{2}}$ so that it contributes less to the scalar curvature than the sphere part does. In anti-de Sitter background ($\eta=-1$), we get the opposite situation; $l^{2}_{ads_{2}}<l^{2}_{s_{2}}$. Only when $l^{2}\rightarrow \infty$, that is when the background gets flat, the sizes of those two factors become coincident.  

\subsection{Non-supersymmetric Attractor Behavior}

Zero Hawking temperature is the property characterizing the extremal black holes [Sec. \ref{seciia}]. Especially it is concerned with the infinite long throat region near the horizon, thus the factorization of the geometry in that region. 

The entropy function of the extremal RNdS black hole provides us with an example exhibiting the non-supersymmetric extremal attractor behavior. Regarding the extremal case, one just needs to focus on the near horizon geometry and the values of the various fields at the horizon. In other words, one may ignore the asymptotic behaviors of the various fields. The attractor behavior is therefore rather concerned with the infinite long throat structure through the factorization of the geometry near the horizon than with the supersymmetry. 

The virtue of the entropy function is that once we are given the action we can compute the extremal black hole entropy even without any knowledge about the full solution. The fact relevant to the determination of the entropy is again that the near horizon geometries of the extremal black holes always involve two-dimenisonal anti-de Sitter space-time. The assumption of the anti-de Sitter space-time as the near horizon geometry is quite universal that it is valid even for the case of higher order gravity. 

\subsection{Higher Order Corrections to the Entropy of RN(A)dS Black Holes}
Being inspired by supergravity/superstring theory, one could consider the higher order corrections to the entropy of Reissner-Nordstr\"{o}m-(Anti-)de Sitter black holes. Let us consider Gauss-Bonnet term, that is, the ghost-free gravitational self-interaction term considered in superstring theory\cite{Zwiebach:1985uq};
\begin{equation}\label{higheraction}
\triangle S= \frac{\alpha}{2\kappa}\int d^{d}x \sqrt{-g} \left(R^{MNPQ}R_{MNPQ}-4 R^{MN}R_{MN}+R^{2} \right). 
\end{equation} 
Here the coefficient of Gauss-Bonnet term $\alpha$ has the dimension of the length squared. 

Inserting the results (\ref{riemann}, \ref{riccitensor}, \ref{scalarcurv}) into Eq. (\ref{higheraction}), we get the following Lagrangian density over AdS$_{2}$;
\begin{eqnarray}\label{lagaddition}
\triangle L&=&\frac{\alpha}{2\kappa}\text{Vol}(S^{d-2}) v_{1}v_{2}^{\frac{d-2}{2}} \left( \frac{1}{v^{2}_{2}}\left(d-4 \right) \left(d-5 \right) - \frac{4 }{v_{1}v_{2}}\right)  \left(d-2 \right) \left( d-3\right). 
\end{eqnarray} 
Especially for $d=4$, only the second term contributes to the Lagrangian density and is constant; $\triangle L=-2\alpha/G_{4}$. Accordingly the entropy function receives a constant correction; $\triangle F=4\pi \alpha/G_{4}$. The total entropy will be 
\begin{eqnarray}\label{q}
S_{\text{tot}}&=&S+\triangle S \nonumber\\
&=& \frac{\pi l^{2}}{6 \,\eta G_{4}}\left(1-\sqrt{1-\frac{3\eta G_{4}}{\pi l^{2}} \left(p^{2}+q^{2} \right)}  \right)+ \frac{4\pi\alpha}{G_{4}}.
\end{eqnarray}

The higher order correction to 4-dimensional black holes results in a constant addition to its entropy. In Eq. (\ref{lagaddition}), we see that Gauss-Bonnet term gives the non-trivial contribution in higher dimensions. 

The above result is very interesting but difficult to understand. It is interesting because Gauss-Bonnet term, though topological in $4$ dimensions, contributes to the entropy, i.e., the degeneracy of the degrees of freedom. Of course there might be some way for the total derivative term to contribute to the entropy via the boundary condition, but then it must involve the ADM mass or the charge. Here, the difficulty resides on its constant nature. 

On the other hand, the same reason gives a clue to view the contribution. Just from the mathematical viewpoint, this constant addition to the entropy looks obvious because Gauss-Bonnet term should be intact, in $4$ dimensions, under the change of the geometric sizes $v_{1,2}$ of AdS and the sphere. Recall that the black hole is believed to be completely characterized by its mass, charge, and angular momentum. Since the contribution is independent of any quantity of these characterizing the black hole, it is reasonable to conclude that the constant does not concern any microscopic degree that constitutes the black hole entropy. We rather employ the viewpoint of Ref. \cite{Myers:1988ze} where similar problem happens and is resolved by regarding the entropy as a relative quantity. One can only use the result (\ref{q})  to determine the entropy of a black hole relative to another in the same class (another cold black hole in the case at hand). 

Another thing to comment is about the signature of the coefficient $\alpha$ of Gauss-Bonnet term. In case it is negative, Gauss-Bonnet term could result in the negative entropy for sufficiently small black hole. In Ref. \cite{Cvetic:2001bk}, it was discussed that the negative entropy may indicate a new type of instability. Indeed the entropy can be negative for either $5$-dimensional Schwarzschild-de Sitter or Schwarzschild-anti-de Sitter black hole (but not both) in Einstein-Gauss-Bonnet gravity. They argue that the appearance of the negative entropy in one type of the black hole may trigger the transition to the other type of the black hole. However, in our case of $4$-dimensional RN(A)dS black hole, for sufficiently negative value of $\alpha$ the entropy can be negative in both anti-de Sitter case and de Sitter case. 

The resolution resides again on the constant nature of the contribution made by Gauss-Bonnet term. Though the negative value of $\alpha$ results in the subtraction of the entropy by a constant amount, we have to view the entropy only in the relative sense because the constant subtraction (or addition) is not involved with any characteristic of the black hole. In higher dimensions, the contribution of Gauss-Bonnet term to the entropy is not constant and there, the signature of $\alpha$ could be very crucial to the viability of the solutions. One can expect some hope from the string theory where the value is positive \cite{Zwiebach:1985uq}.

\acknowledgments
We thank Ashoke Sen for helpful discussions which inspired this work very much.
This work is supported by the SRC program of KOSEF through CQUeST with grant number R11-2005-021.

\end{document}